\documentclass[conference]{IEEEtran}
\IEEEoverridecommandlockouts
\usepackage{amsmath,amsfonts}
\usepackage{algorithmic}
\usepackage{algorithm}
\usepackage{array}
\usepackage[caption=false,font=normalsize,labelfont=sf,textfont=sf]{subfig}
\usepackage{textcomp}
\usepackage{stfloats}
\usepackage{url}
\usepackage{verbatim}
\usepackage{graphicx}
\usepackage{cite}
\usepackage{xcolor}
\usepackage{mathtools}
\usepackage{amsmath, amsfonts, amssymb, amsthm}
\usepackage{setspace}
\usepackage{ulem}
\usepackage[numbers,sort&compress]{natbib}
\usepackage{booktabs}
\usepackage{siunitx}
\usepackage{makecell}
\usepackage{multirow}
\usepackage{amsmath}
\usepackage{tabularx}

\usepackage{soul}
\usepackage{xcolor}
\begin{document}
\title{UAV-based Energy-Efficient Data Collection in Smart Grids with ISAC QoS Guarantees}
\author{
\IEEEauthorblockN{Yibin Xie\IEEEauthorrefmark{1},
Jin Zhao\IEEEauthorrefmark{1},
Indrakshi Dey\IEEEauthorrefmark{2},
Nicola Marchetti\IEEEauthorrefmark{1}\thanks{This work is supported by the US-Ireland R\&D Partnership Programme Project "Resilient Networks" under Grant RI-SFI-23/US/3924, the EU MSCA Project "COALESCE" under Grant Number 101130739, Research Ireland Grant 13/RC/2077\_P2, and the China Scholarship Council-Trinity College Dublin Joint Scholarship Programme (Grant No.202406690018).}}
\vspace{1mm}

\IEEEauthorblockA{\IEEEauthorrefmark{1}Department of Electronic and Electrical Engineering, Trinity College Dublin, Ireland}

\IEEEauthorblockA{\IEEEauthorrefmark{2}Walton Institute, South East Technological University, Waterford, Ireland}

\IEEEauthorblockA{Emails: yixie@tcd.ie, zhaoj6@tcd.ie, indrakshi.dey@waltoninstitute.ie, nicola.marchetti@tcd.ie}
\vspace{-12mm}
}

\maketitle
\thispagestyle{empty}
\begin{abstract}
   Dynamic line rating (DLR) is a methodology that requires timely monitoring data to determine the real-time ampacity of power lines.
   However, DLR monitoring devices (MD) are vulnerable to connectivity disruptions, leading to missing or delayed data.
   Although unmanned aerial vehicles (UAV) can enable resilient data collection from MD, their limited onboard energy challenges timely monitoring over extended transmission corridors with flight hazards.
   This paper proposes a cooperative UAV-based data collection framework with integrated sensing and communication (ISAC) to support timely DLR updates.  
   In this framework, ISAC is employed to maintain the sensing and communication quality required for safe and cooperative UAV data collection.
   Accordingly, a joint energy minimization problem is formulated over UAV trajectories and collection scheduling under ISAC constraints. To solve it, a hybrid algorithm combining deep reinforcement learning (DRL) and semidefinite relaxation (SDR) is proposed, where DRL optimizes the trajectory and collection scheduling, while SDR is used to handle the non-convex ISAC constraints.
   Simulation results show that the proposed scheme reduces energy consumption by up to 34.6\% compared with offline benchmarks and by about 2.2\% compared with the separated sensing-and-communication baseline, while satisfying the minute-level timescale requirement of DLR.
\end{abstract}


\vspace{-2mm}
\section{Introduction}
Smart grids are cyber-physical systems that integrate power infrastructure with advanced communication and information technologies for reliable and efficient power delivery. 
As transmission lines form the backbone of power delivery, continuous line monitoring is indispensable for operational safety and asset utilization.
Building on such monitoring, dynamic line rating (DLR) has emerged as a key technology for improving line utilization, since real-time thermal limits derived from environmental conditions (e.g., temperature, wind speed) and line characteristics (e.g., conductor sag) can substantially increase ampacity compared with static ratings \cite{karimi2018dynamic}.

However, the effectiveness of DLR relies on timely data acquisition through Internet of Things (IoT) monitoring devices (MD) and reliable delivery via ground-based communication infrastructure along transmission corridors \cite{wydra2019time}. Under extreme events, such infrastructure may fail due to power outages, physical damage, or signal disruption, creating information islands. Operators are then forced to fall back on conservative static ratings or stale ratings, which can either underutilize transmission capacity or underestimate thermal stress and risk conductor overheating.

Unmanned aerial vehicles (UAV) have recently become an auxiliary enabler for IoT data collection when terrestrial links are limited \cite{10065524}, and can therefore complement smart grid monitoring during communication disruptions. Existing UAV applications in power systems mostly target line inspection and fault detection along fixed corridor routes \cite{10603401}, whereas data collection requires UAVs to visit geographically dispersed MDs under energy constraints, making trajectory planning critical. Moreover, DLR updates are typically required at minute-level timescales \cite{poli2019possible}, so a single UAV may be insufficient to serve dispersed MDs within the time budget, motivating multiple cooperative UAVs for timely and energy-efficient collection.

Transmission corridors also create challenging low-altitude airspaces, where hazards such as vegetation, terrain, and especially power lines threaten UAV operations through collision and electromagnetic interference \cite{dianovsky2023electromagnetic}. Millimeter-wave (mmWave) radar sensing is well suited to detecting such thin metallic structures owing to its high resolution \cite{malle2021survey}. Furthermore, integrated sensing and communication (ISAC) allows radar and communication links to share the same hardware and spectrum \cite{10769423}, improving spectrum efficiency and reducing UAV payload and energy overhead. 
This combination makes mmWave-enabled ISAC attractive for smart grid UAVs, where mmWave sensing can perceive corridor hazards while wireless links support inter-UAV cooperation. To the best of our knowledge, however, energy-efficient UAV-based data collection under joint communication and sensing quality of service (QoS) requirements has not yet been investigated.

Motivated by the above, this paper investigates an ISAC-enabled UAV data collection framework for DLR in smart grids. The main contributions are summarized as follows.
\begin{itemize}
\item We propose an ISAC-enabled UAV data collection framework for DLR monitoring, where ISAC QoS is characterized by inter-UAV communication signal-to-interference-plus-noise (SINR) and transmit beampattern (TBP) gain.

\item We formulate an energy minimization problem that jointly optimizes UAV trajectories and MD collection scheduling under ISAC QoS constraints.

\item We develop a hybrid deep reinforcement learning and semidefinite relaxation (DRL-SDR) solution that combines DRL-based cooperative optimization with SDR-based ISAC transmit design. Simulation results verify its effectiveness in energy efficiency and collection performance.
\end{itemize}

\vspace{-6mm}
\section{System model}
As shown in Fig. 1, a UAV-assisted grid monitoring system for DLR is considered.
A set of $I$ single antenna MDs indexed by $\mathcal I = \{ 1, 2, \cdots, I \}$ is distributed along the transmission lines and nearby areas to monitor line and environmental conditions.
Due to limited ground-based communication, the grid control center cannot timely receive data from the MDs, and UAVs are deployed for data collection.
Let $\mathcal{M} = \{1, \ldots, M\}$ denote the index set of UAVs, and let the set of UAVs be given by $\mathcal{U} = \{ U_m \mid m \in \mathcal{M} \}$.
Each UAV is equipped with a single omnidirectional antenna for communication with MD, and a vertically placed uniform linear array (ULA) \cite{10879807} with $L$ elements for inter-UAV communication and radar probing to support safe flight along transmission line corridors.
The two antenna systems operate on orthogonal frequency resources.
During the mission, the UAVs depart from an initial station, form a decentralized inter-UAV network, collect data from distributed MDs, and forward the data to the grid operator after reaching the final position.
\vspace{-3mm}
\begin{figure}[!ht]
   \centering
   \setlength{\abovecaptionskip}{2pt}
   \setlength{\belowcaptionskip}{-4pt}
   \includegraphics[width=3.4in]{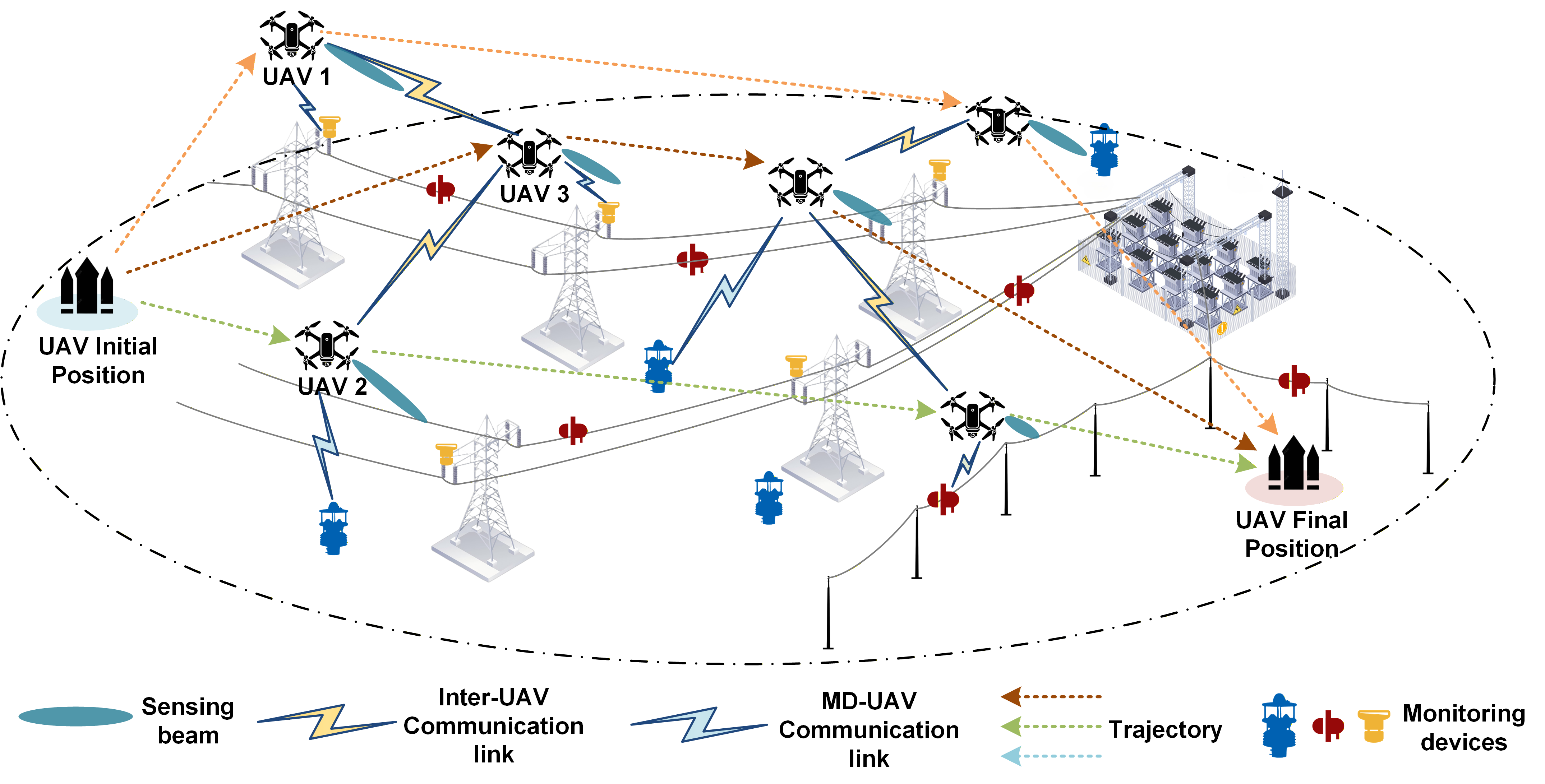}
   \caption{Overview of the system model.}
   \vspace{-2mm}
   \label{fig1}
\end{figure}
\vspace{-5mm}

\subsection{UAV mobility model}
During the mission, we consider a frame structure with time slot length $\tau$.
Each slot is indexed by $t$, and the slot set is denoted as $\mathcal T = \{t \mid 1 \leq t \leq T \}$.
Each UAV is set to fly at a fixed altitude $z_0$ and the position of UAVs under a three-dimensional Cartesian coordinate system in each slot can be denoted as $\boldsymbol{q}_{m}[t] = [x_{m}[t], y_{m}[t], z_0]^T$,
and the overall flight trajectory can be denoted by $\boldsymbol{q}_m = \{\boldsymbol{q}_{m}[1], \cdots, \boldsymbol{q}_{m}[T]\}$.
The UAVs are required to depart from the initial location and reach the final location, i.e,
\begingroup
\fontsize{9.6pt}{11pt}\selectfont
\begin{equation}
 \boldsymbol{q}_{m}[1] = [x_0, y_0, z_0]^T, \quad
 \boldsymbol{q}_{m}[T] = [x_1, y_1, z_0]^T, \
 \forall m \in \mathcal{M}.
\end{equation}
\endgroup
Moreover, each UAV is assumed to either hover or move at a constant speed $V_{\mathrm{fixed}}$ along a controlled heading direction.

\vspace{-3mm}
\subsection{UAV data collection model}
To indicate whether $U_m$ collects data from the $i$-th MD at each time slot, we introduce the following binary variable:
\begingroup
\fontsize{9.6pt}{11pt}\selectfont
\begin{equation}
    \begin{array}{ll}
        b_{m,i}[t] = \begin{cases}
        1 \ , &\mbox{if $U_m$ collect data from $i$-th MD at slot $t$,} \\
        0 \ , &\mbox{otherwise.}\\
        \end{cases}
	\end{array}	
\end{equation}
\endgroup
The locations of the MD in the three-dimensional Cartesian coordinate system are denoted by $\boldsymbol{u} = \{ \boldsymbol{u}_1, \boldsymbol{u}_2, \cdots, \boldsymbol{u}_I \}$, where $\boldsymbol{u}_i = [x_i, y_i, z_i]^T$
represents the location of $i$-th MD.
Since the line-of-sight (LoS) component is dominant in the UAV-MD communication links, and considering that MDs are installed at varying altitudes, we adopt a probabilistic LoS channel model \cite{8663615} to accurately capture the link characteristics.
The probability of the LoS part can be computed as:
\begingroup
\fontsize{9.6pt}{10pt}\selectfont
\begin{equation}
    P_{m,i}^{LoS}[t] = \frac{1}{1 + C \, \text{exp}(-D\,(\, \theta_{m,i}[t] - C\,))}.
\end{equation}
\endgroup
where $C$ and $D$ are the parameters depending on the propagation environment, $\theta_{m,i}[t]$ is the elevation angle between $U_m$ and MD $i$ in slot $t$.
Accordingly, the expected channel can be given by
\begingroup
\fontsize{9.6pt}{10pt}\selectfont
\begin{equation}
    h_{m,i}[t] = P_{m,i}^{LoS}[t]\, \frac{\sqrt{\beta_i}}{d_{m,i}[t]} +
    (1-P_{m,i}^{LoS}[t])\, \kappa \frac{\sqrt{\beta_i}}{d_{m,i}[t]}.
\end{equation}
\endgroup
where $\beta_i$ denotes the signal path loss at reference distance $d_0 = 1$m, $d_{m,i}[t]=\left\|\boldsymbol q_{m}[t]-\boldsymbol u_i\right\|_2$ is the distance between $U_m$ and MD $i$, and $\kappa$ denotes the attenuation factor due to the non-line-of-sight (NLoS) propagation.

During flight, multiple MDs may be located within the coverage area of a certain UAV.
To avoid interference from simultaneous transmission, we assume that in each time slot, each UAV schedules at most one MD for data uploading.
Consequently, the uplink transmission suffers only inter-cell interference from MDs served by other UAVs.
The corresponding received SINR can be expressed as
\begingroup
\fontsize{9.6pt}{11pt}\selectfont
\begin{equation}
    \gamma_{m,i}[t] = \frac{p_i \, h_{m,i}[t] \, b_{m,i}[t]}{\sum\limits_{m' \in \mathcal M \setminus \{m\}}\sum\limits_{i' \in \mathcal I} b_{m',i'}[t] ~ p_{i'} \, h_{m,i'}[t] + \sigma_{\text{MD}}^2}.
\end{equation}
\endgroup
where $p_i$ is the transmit power of $i$-th MD, $\sigma_{\text{MD}}^2$ represents the effective noise power at the receiver.
Since DLR monitoring data packets are typically lightweight \cite{wydra2019time}, data collection is regarded as successful once the received SINR at the UAV exceeds the communication threshold \scalebox{0.98}{$\gamma_{\text{th}}^{\text{MD}}$}.

\subsection{Inter-UAV communication and sensing QoS model}
For tractability, we model the inter-UAV communication topology as a predefined and time-invariant chain, represented by the directed edge set $\mathcal{E}$. 
Each directed edge $(m,m')\in\mathcal{E}$ represents a scheduled communication link from $U_m$ to $U_{m'}$.
Within each time slot, the links in $\mathcal{E}$ are activated sequentially in a time-division manner, allowing each active UAV to communicate with its designated neighbour without network-level interference.

The ULA mounted on each UAV is shared by the communication and radar systems. 
When link $(m,m')$ is activated, $U_m$ transmits information to $U_{m'}$ while simultaneously transmitting dedicated sensing signals toward predefined directions for hazard awareness in transmission corridors.
Thus, the transmit signal of $U_m$ is a superposition of information signals and radar waveforms \cite{10879807}, given by
\begingroup
\fontsize{9.6pt}{10pt}\selectfont
\begin{equation}
\boldsymbol{x}_m[t] = \boldsymbol{w}_m^{c}[t] c_m[t] + \boldsymbol{s}_m[t].
\end{equation}
\endgroup
Here, $c_m[t]$ is the information signal from $U_m$ to its desired receiver at slot $t$, and \scalebox{0.9}{$\boldsymbol{w}_m^{c}[t] \in \mathbb{C}^{L \times 1}$} represents the beamforming vector for the information signal. $\boldsymbol{s}_m[t] \in \mathbb{C}^{L \times 1}$ represents the dedicated sensing signal vector. 
Both $c_m[t]$ and $\boldsymbol{s}_m[t]$ are assumed to be zero-mean and mutually independent. 
Furthermore, $c_m[t]$ is normalized to unit power, and the spatial covariance matrix of the sensing signal is defined as $\boldsymbol{R}^s_m[t] = \mathbb{E}[\boldsymbol{s}_m[t]\boldsymbol{s}_m[t]^H] \succeq \boldsymbol{0}$.
Then, the transmit beampattern (TBP) gain \cite{10769423} in an angular direction $\phi$ can be expressed by
\begingroup
\fontsize{9.6pt}{10pt}\selectfont
\begin{equation}
P^{\text{TBP}}_m(\phi)[t]= \boldsymbol{a}^H(\phi) \left( \boldsymbol{w}_m^{c}[t] (\boldsymbol{w}_m^{c}[t])^H + \boldsymbol{R}^s_m[t] \right) \boldsymbol{a}(\phi),
\end{equation}
\endgroup
where $\boldsymbol{a}(\phi) = [1, e^{j\pi\sin\phi}, \cdots, e^{j\pi (L-1) \sin\phi} ]^{\mathsf T}$ is the steering vector.
To maintain the sensing capability for hazard awareness, the TBP gain is required to exceed a given threshold $\Gamma$ at each of the $K$ discrete angular directions $\Phi=\{\phi_1,\ldots,\phi_K\}$.

Each inter-UAV link is modeled using a Rician fading channel \cite{10791445}, given by:
\begingroup
\fontsize{9.6pt}{10pt}\selectfont
\begin{equation}
   \boldsymbol{H}_{m',m}[t] = \sqrt{\frac{K}{K+1}}\boldsymbol{H}_{m',m}^{LoS}[t] + \sqrt{\frac{1}{K+1}} \boldsymbol{H}_{m',m}^{NLoS}[t],
\end{equation}
\endgroup
where $K$ is the Rician factor for UAV communications, $\boldsymbol{H}_{m',m}^{LoS}[t]$ denotes the LoS channel from $U_{m}$ to $U_{m'}$.
Since UAVs operate at the same altitude and employ vertical ULAs, the angles of arrival and departure are both zero. Thus, $\boldsymbol{H}_{m',m}^{LoS}[t]$ can be modeled as \scalebox{0.96}{$\boldsymbol{H}_{m',m}^{LoS}[t]= \frac{\sqrt{\beta_m}}{d_{m',m}[t]} \boldsymbol{1}_{L} \boldsymbol{1}_{L}^H$}, where \scalebox{0.95}{$d_{m',m}[t] = \left\|\boldsymbol q_{m}[t]-\boldsymbol q_{m'}[t]\right\|$} is the distance between $U_m$ and $U_{m'}$,  \scalebox{0.95}{$\boldsymbol{1}_{L}$} and \scalebox{0.95}{$\boldsymbol{1}_{L}^H$} are the steering vectors. 
Additionally, \scalebox{0.96}{$\boldsymbol{H}_{m',m}^{\mathrm{NLoS}}[t] \in \mathbb{C}^{L \times L}$} is the NLoS channel matrix with i.i.d. entries following $\mathcal{CN}(0,1)$.
Under the time-division transmission assumption, simultaneous inter-UAV link interference is avoided.
Moreover, since ISAC signaling and data collection flow are allocated on orthogonal frequency resources, they can be executed in parallel without mutual interference.
Therefore, the interference at the receiver mainly comes from the sensing signal transmitted by $U_m$.
Accordingly, the received SINR from $U_{m}$ to $U_{m'}$ in a slot can be computed by
\begingroup
\fontsize{9.6pt}{10pt}\selectfont
\begin{equation}
    \gamma_{m',m}[t] = \frac{|\boldsymbol{f}_{m'}^H \boldsymbol{H}_{m',m}[t] \boldsymbol{w}_{m}^c[t]|^2}{\boldsymbol{f}_{m'}^H \boldsymbol{H}_{m',m}[t] \boldsymbol{R}^s_{m}[t] \boldsymbol{H}_{m',m}^H[t] \boldsymbol{f}_{m'} + \sigma_U^2},
\end{equation}
\endgroup
where $\boldsymbol{f}_{m'}^H \in \mathbb{C}^{1 \times L}$ denotes the receive beamforming vector, which can be treated as a constant since all UAVs fly at the same altitude. $\sigma_{\text{U}}^2$ represents the noise power.
The communication QoS for each link must satisfy a threshold, i.e., $\gamma_{m',m}[t] \geq \gamma_{\text{th}}^{\text{U}}$. 

\vspace{-1mm}
\subsection{Energy consumption model}
Since the transmit power for communication and sensing is much lower than the propulsion power required for UAV flight and hovering \cite{8663615}, we only consider the propulsion energy consumption in this work.
Let $E_m[t]$ denote the whole energy consumed by $U_m$ in each time slot, which is given by
\begingroup
\fontsize{9.6pt}{10pt}\selectfont
\begin{equation}
    E_m[t] = \big[\mu_m[t] P^f_m[t] + (1-\mu_m[t]) P^h_m[t]\big]\tau .
\end{equation}
\endgroup
Here, $\mu_m[t]$ is the flight variable of $U_m$, where if $U_m$ is in flying status in time slot $t$, $\mu_m[t] = 1$, otherwise, $\mu_m[t] = 0$.
Under the fixed flying speed $V_{\mathrm{fixed}}$, the flying propulsion power $P^f_m[t]$ remains constant during flight and consists of blade profile, induced, and fuselage parasite power components \cite{8663615}.
For hovering, the corresponding power consumption is $P^h_m[t]=P_b+P_r$, where $P_b$ and $P_r$ are constant blade profile power and induced power, respectively.

\vspace{-1mm}
\section{Problem Formulation}
The main objective of this work is to minimize the UAV energy consumption by jointly optimizing data collection scheduling and UAV trajectories, while ensuring inter-UAV SINR and TBP gain requirements through feasible communication beamforming and sensing covariance matrix design.
Let $\boldsymbol{\Psi} \triangleq \{\boldsymbol{q}, \boldsymbol{b}, \boldsymbol{W}^c, \boldsymbol{R}^s\}$ denote the set of all variables, where $\boldsymbol{b} = \{b_{m,i}[t], \, \forall m, i, t\}$, $\boldsymbol{q} = \{\boldsymbol{q}_m, \, \forall m\}$, $\boldsymbol{W}^c = \{\boldsymbol{w}^c_m[t], \, \forall m, t\}$, and $\boldsymbol{R}^s = \{\boldsymbol{R}^s_m[t], \, \forall m, t\}$.
Therefore, the optimization problem is formulated as:
\begingroup
\fontsize{9.5pt}{11pt}\selectfont
\allowdisplaybreaks
\begin{align}
\hspace{-0.5em} \mathcal P1: ~\min_{\boldsymbol{\Psi}} & ~\sum_{t\in\mathcal{T}} \sum_{m\in\mathcal{M}} E_m[t] \label{Problem}\\
\text {s.t.} & \quad \text{(1)} - \text{(10)}, \tag{\ref{Problem}a} \\
& \sum\limits_{m \in \mathcal M}b_{m,i}[t] \leq 1, \ \forall i \in \mathcal I, t \in \mathcal T, \tag{\ref{Problem}b} \\
& \sum_{t\in\mathcal{T}} \sum_{m\in\mathcal{M}} b_{m,i}[t] \geq 1, \ \forall i \in \mathcal I, \tag{\ref{Problem}c} \\
&~ ||\boldsymbol{w}^c_m[t]||^2 + \mathrm{Tr}(\boldsymbol{R}^s_m[t]) \leq p_{\max},
~\forall m \in \mathcal M, t \in \mathcal T, \tag{\ref{Problem}d}  \\
&~ \boldsymbol{R}^s_m[t] \succeq \boldsymbol{0}, \tag{\ref{Problem}e} \forall m \in \mathcal M, t \in \mathcal T, \notag \\
&~ P^{\mathrm{TBP}}_m(\phi_k)[t] \geq \Gamma, \; \forall \phi_k \in \Phi, m \in \mathcal M, t \in \mathcal T, \tag{\ref{Problem}f} \\
&~ d_{m',m}[t] \geq d_{\min}, \ \forall m,m'\in\mathcal{M},\; m\neq m',\; t\in\mathcal{T}, \tag{\ref{Problem}g} \\
&~ \gamma_{m,i}[t] \geq \gamma_{\text{th}}^{\text{MD}} b_{m,i}[t], \ \forall m \in \mathcal{M}, t \in \mathcal T, \tag{\ref{Problem}h} \\
&~ \gamma_{m',m}[t] \geq \gamma_{\text{th}}^{\text{U}}, \ \forall (m, m') \in \mathcal{E}, t \in \mathcal T \tag{\ref{Problem}i}
\end{align}
\endgroup
where (\ref{Problem}{b}) constrains that each MD can only upload its data to one dedicated UAV per slot;
(\ref{Problem}{c}) ensures that every MD is successfully served at least once;
(\ref{Problem}{d}) ensures that the transmit power of each UAV is less than the maximum power $p_{\max}$;
(\ref{Problem}{e}) makes sure the covariance matrix of sensing signal is positive semi-definite;
(\ref{Problem}{f}) represents the TBP gain in direction $\phi_{k}$ should exceeds a threshold $\Gamma$.
(\ref{Problem}{g}) enforces a minimum safety distance $d_{\min}$ between any pair of UAVs.
(\ref{Problem}{h}) and (\ref{Problem}{i}) guarantee the minimum SINR requirement of MD-UAV and inter-UAV communication links, respectively.

Problem $\mathcal{P}1$ is a mixed-integer non-linear programming (MINLP) problem with tightly coupled variables $\boldsymbol{q}, \boldsymbol{b}, \boldsymbol{W}^c, \boldsymbol{R}^s$. 
Specifically, trajectory $\boldsymbol{q}$ determines the MD uplink quality and feasible collection decisions, while the scheduling decisions $\boldsymbol{b}$ influence subsequent UAV movements. Meanwhile, beamforming vector $\boldsymbol{W}^c$ and sensing covariance matrix $\boldsymbol{R}^s$ are coupled through the shared transmit power budget, making the inter-UAV SINR and sensing TBP gain constraints interdependent as well. 
Therefore, directly solving $\mathcal{P}1$ is computationally intractable. To address this, we propose a hierarchical decoupling framework that decomposes the original problem into two tractable subproblems: an upper-layer DRL-based trajectory and collection optimization, in which DRL agents are trained and employed to jointly optimize $\boldsymbol{q}$ and $\boldsymbol{b}$; and lower-layer ISAC transmit design, in which, given the UAV positions generated by the DRL agents at each slot, the \scalebox{0.98}{$\boldsymbol{w}^c_m[t]$} and \scalebox{0.98}{$\boldsymbol{R}^s_m[t]$} are optimized via SDR to check the feasibility of the inter-UAV SINR and TBP gain constraints.

\vspace{-2mm}
\section{Joint Trajectory and Collection Scheduling Design with ISAC QoS constraints}
\subsection{DRL-based trajectory \& collection scheduling optimization}
In this section, we propose a multi-agent deep reinforcement learning-based method to jointly optimize UAV trajectory and collection scheduling. 
Specifically, we first formulate the joint optimization problem as a Markov decision process (MDP) and propose a multi-agent proximal policy optimization (MAPPO)-based solution. 
The centralized training and decentralized execution (CTDE) paradigm \cite{10065524} is adopted, where global state information is utilized during training, whereas each UAV executes its policy independently based only on local observations. Therefore, the proposed DRL-based algorithm achieves cooperative trajectory control and collection scheduling while supporting practical distributed deployment.
The MDP components are defined as follows:

\textit{1) Observation:}
Each UAV's observation $\mathbf{o}_m^t$ contains its forward direction $\psi_m[t]$, current location $q_m[t]$, residual energy $E_m^r[t] = E_{\text{total}} - \sum_{n=1}^t E_{m}[n]$, the relative distance to MDs, and landing point. With the help of inter-UAV communication links, each UAV is also aware of the relative location of other UAVs and the current collection status vector $\boldsymbol{c}[t] \in \{0, 1\}^I$, where $c_i[t]=1$ indicates that the $i$-th MD's data has been collected by any UAV by slot $t$.

\textit{2) State Space:} Under CTDE, the centralized critic state $\mathbf{s}^t$ consists of the joint observations of all UAVs together with the global MD-location information and collection status.

\textit{3) Action Space:}
At each time slot, UAV observes its local state $\mathbf{o}_m^t$ and selects a compound action $a_m^t=(b_m[t],\psi_m[t], v_m[t])$, which consists of an MD scheduling decision $b_m[t]$, the heading angle $\psi_m[t]$, and the flying speed $v_m[t]$. Here, the flying speed is restricted to $v_m[t] \in \{0, V_{\text{fixed}}\}$, meaning the UAV can only choose to hover or fly at a predetermined fixed speed.
To avoid infeasible or redundant scheduling decisions, an action mask is applied to the MD selection process, excluding MDs that have already been served or are currently scheduled by other UAVs.

\textit{4) State transition:}
Given the joint action $\mathbf{a}^t=\{(b_m[t],\psi_m[t], v_m[t]),\,m\in\mathcal{M}\}$ at slot $t$, the system state evolves according to the UAV motion, MD collection status, and energy consumption. Specifically, the position of $U_m$ is updated as
\begingroup
\fontsize{9.5pt}{10pt}\selectfont
\begin{equation}
\boldsymbol{q}_m[t+1] = \boldsymbol{q}_m[t] + v_m[t]\tau
\begin{bmatrix}
\cos\psi_m[t]\\
\sin\psi_m[t] \\
0
\end{bmatrix},
\quad \forall m\in\mathcal{M}.
\end{equation}
\endgroup
The collection status of MD $i$ is updated as
\begingroup
\fontsize{9.5pt}{10pt}\selectfont
\begin{equation}
    c_i[t+1] = \min \left( 1, \ c_i[t] + \sum_{m=1}^{M} b_{m,i}[t] \right), \ \forall i \in \mathcal I.
\end{equation}
\endgroup
By updating all $I$ MDs, the collection status vector $\mathbf{c}[t]$ will be updated.
Finally, the residual energy of $U_m$ evolves as \scalebox{0.97}{$E_m^r[t+1] = E_m^r[t] - E_m[t], \ \forall m\in\mathcal{M}$}.

\textit{5) Reward:} Under the CTDE framework, the shared reward for all UAV agents is defined as $r^t = R_{c}[t] + R_{q}[t] + P_{e}[t] + P_{d}[t]$, where $R_{c}[t]$ denotes the total collection reward of all UAVs, $R_{q}[t]$ represents the ISAC QoS reward evaluated from the SDR feasibility check, $P_{e}[t]$ represents the penalty caused by energy consumption,  and $P_{d}[t]$ denotes the penalty for violating the minimum inter-UAV distance constraint.

The MAPPO framework adopts an actor-critic structure, where the actor network is optimized using proximal policy optimization.
For simplicity, we omit the agent index in the following formulations.
Let $\pi_{\nu}$ and $\pi_{\nu_{\mathrm{old}}}$ denote the new and old policies, respectively. 
For each sampled transition, the probability ratio is defined as $\rho^t(\nu)=\pi_{\nu}(\mathbf{a}^t|\mathbf{o}^t)/\pi_{\nu_{\mathrm{old}}}(\mathbf{a}^t|\mathbf{o}^t)$.
The actor is then updated by maximizing
\begingroup
\fontsize{9.5pt}{10pt}\selectfont
\begin{equation}\label{au}
J_{a}(\nu)=\mathbb{E}\!\left[
\min\!\left(
\rho^t(\nu)\hat{A}^t,\,
\mathrm{clip}(\rho^t(\nu),1-\epsilon,1+\epsilon)\hat{A}^t
\right)\right].
\end{equation}
\endgroup
The clipping function $\mathrm{clip}(\cdot)$ constrains the probability ratio $\rho^t(\nu)$ within $[1-\epsilon,1+\epsilon]$, preventing destructive policy oscillation. 
The advantage estimate $\hat{A}^t$ is computed by generalized advantage estimation (GAE) \cite{10097711}, i.e., 
$\hat{A}^t=\sum_{n=0}^{T-t}(\eta\lambda)^n \delta^{t+n}$, 
where $\eta$ is the discount factor, $\lambda$ is the GAE parameter, and $\delta^t$ denotes the temporal-difference residual. 

The critic network is updated by minimizing the mean squared error (MSE) between the predicted state values and the target values. The MSE for the critic network is
\begingroup
\fontsize{9.5pt}{10pt}\selectfont
\begin{equation}\label{cu}
    J_{c}(\omega) = \mathbb{E} \left[ \left( V(\mathbf{s}^t) - V_{\text{target}}(\mathbf{s}^t) \right)^2 \right],
\end{equation}
\endgroup
where $V(\mathbf{s}^t)$ is the state value estimated by the current critic network parameterized by $\omega$. The target value $V_{\text{target}}(\mathbf{s}^t)$ can be obtained by $V_{\text{target}}(\mathbf{s}^t) = \hat{A}^t + V_{\text{old}}(\mathbf{s}^t)$, where $V_{\text{old}}(\mathbf{s}^t)$ is the value predicted by the old critic network used during the trajectory sampling.

\vspace{-3.5mm}
\subsection{SDR-based ISAC transmit design}
Once the UAV positions are obtained in each slot, the ISAC sensing and communication requirements are evaluated by solving the following feasibility-check problem for the transmitter $U_m$ of each scheduled directed link $(m,m')\in\mathcal{E}$.
\begingroup
\fontsize{9.6pt}{11pt}\selectfont
\allowdisplaybreaks
\begin{align}
\hspace{-1em} \mathcal P2: \text{find} & \quad \boldsymbol{w}^c_m[t], \boldsymbol{R}_m^s[t]  \label{Problem1}\\
\text{s.t.} & \:  P^{\mathrm{TBP}}_m(\phi_{k})[t] \geq \Gamma, 
~ \forall \; \phi_k \in \Phi, \tag{\ref{Problem1}a} \\
&~ \gamma_{m',m}[t] \geq \gamma_{\text{th}}^{\text{U}}, \tag{\ref{Problem1}{b}} \\
&~ ||\boldsymbol{w}^c_m[t]||^2 + \operatorname{Tr}(\boldsymbol{R}^s_m[t]) \leq p_{\max}, \tag{\ref{Problem1}{c}} \\
&~ \boldsymbol{R}^s_m[t] \succeq \boldsymbol{0}. \tag{\ref{Problem1}{d}} \notag
\end{align}
\endgroup
We define the covariance matrix of the communication beamforming vector as $\boldsymbol{R}^c_m[t] = \boldsymbol{w}^c_m[t] {\boldsymbol{w}^c_m[t]}^H$. 
Therefore, the total transmit covariance matrix is defined as: $\boldsymbol{R}_m[t] = \boldsymbol{R}^c_m[t] + \boldsymbol{R}^s_m[t]$. We further define the effective channel matrix as $\tilde{\boldsymbol H}_{m',m}[t] = \boldsymbol H_{m',m}[t]^{H}\boldsymbol f_{m'}\boldsymbol f_{m'}^{H}\boldsymbol H_{m',m}[t]$. By using $|\boldsymbol x|^{2}= \operatorname{Tr}(\boldsymbol x \boldsymbol x^{H})$, it follows that
$|\boldsymbol f_{m'}^{H}\boldsymbol H_{m',m}[t]\boldsymbol w^{c}_{m}[t]|^{2} = \operatorname{Tr}\ ((\boldsymbol f_{m'}^{H}\boldsymbol H_{m',m}[t]\boldsymbol w^{c}_{m}[t])(\boldsymbol f_{m'}^{H}\boldsymbol H_{m',m}[t]\boldsymbol w^{c}_{m}[t])^{H}) = \operatorname{Tr}(\boldsymbol R^{c}_{m}[t] \, \boldsymbol{\tilde{H}}_{m',m}[t])$. 
Then, the communication SINR between UAVs can be re-written as
$\gamma_{m',m}[t] = \frac{\operatorname{Tr}(\boldsymbol R^{c}_{m}[t] \, \boldsymbol{\tilde{H}}_{m',m}[t])} {\operatorname{Tr}(\boldsymbol R^{s}_{m}[t] \, \boldsymbol{\tilde{H}}_{m',m}[t]) + \sigma_{U}^2}.$
Thus, the beamforming optimization subproblem ($\mathcal P2$) can be reformulated as:

\vspace{-1mm}
\begingroup
\fontsize{9.6pt}{11pt}\selectfont
\allowdisplaybreaks
\begin{align}
\hspace*{-0.5em} \mathcal P2.1: \text{find} & \quad \boldsymbol{R}^c_m[t], \, \boldsymbol{R}^s_m[t] \label{Problem2}\\
\text{s.t.} & \: \boldsymbol{a}^H(\phi_{k}) (\boldsymbol{R}^c_m[t] + \boldsymbol{R}^s_m[t]) \boldsymbol{a}(\phi_{k}) \geq \Gamma, \tag{\ref{Problem2}{a}} \\
&~ \; \forall \phi_k \in \Phi, \notag \\
&~ [\operatorname{Tr}(\boldsymbol R^{s}_{m,t} \boldsymbol{\tilde{H}}_{m',m}[t]) + \sigma_{U}^2] \, \gamma_{\text{th}}^{\text{U}} \leq \notag \\
&~ \operatorname{Tr}(\boldsymbol R^{c}_{m}[t] \, \boldsymbol{\tilde{H}}_{m',m}[t]), \tag{\ref{Problem2}{b}} \\
&~ \operatorname{Tr}(\boldsymbol{R}^c_m[t]) + \operatorname{Tr}(\boldsymbol{R}^s_m[t]) \leq p_{\max}, \tag{\ref{Problem2}{c}} \\
&~ \operatorname{rank} (\boldsymbol{R}^c_m[t]) = 1, \, \tag{\ref{Problem2}{d}} \\
&~ \boldsymbol{R}^c_m[t] \succeq \boldsymbol{0}, \,  \tag{\ref{Problem2}{e}} \\
&~ \boldsymbol{R}^s_m[t] \succeq \boldsymbol{0}, \, \tag{\ref{Problem2}{f}}
\end{align}
\endgroup
The constraint (\ref{Problem2}{d}) enforces a single communication beam, while (\ref{Problem2}{e}) ensures the transmit power is non-negative.
Problem $\mathcal{P}2.1$ is non-convex only due to the rank-1 constraint on $\boldsymbol{R}^c_m[t]$ in (\ref{Problem2}{d}).
We therefore apply SDR by dropping (\ref{Problem2}{d}), which converts $\mathcal{P}2.1$ into a convex semidefinite program (SDP).
The SDR solution provides a lower bound to problem  $\mathcal{P}2.1$, which becomes tight when the obtained covariance matrix is exactly or nearly rank-one.
The relaxed SDP can then be efficiently solved by standard convex solvers like CVX. Finally, the proposed DRL-SDR algorithm for trajectory and collection scheduling optimization with ISAC QoS requirements is summarized in Algorithm 1\footnote{At each slot, the upper-layer DRL policy requires $\mathcal{O}(M D_{\rm actor})$ operations for $M$ UAV agents, where $D_{\rm actor}\triangleq \sum_{g=1}^{G} d_{g-1}^a d_g^a$ denotes the cost of one actor-network forward pass, with $d_{g-1}^{a}$ and $d_{g}^{a}$ denoting the input and output dimensions of the $g$-th layer. The lower-layer SDR is dominated by the $L\times L$ semidefinite matrix variable, with worst-case interior-point complexity $\mathcal{O}(L^6)$ per link. Hence, the overall per-slot complexity is approximately $\mathcal{O}(M D_{\rm actor}+|\mathcal{E}|L^6)$, where $|\mathcal{E}|$ denotes the number of links.}.

\vspace{-2mm}
\begin{algorithm}[!htbp]
\caption{The proposed DRL-SDR Algorithm}\label{alg1}
{\fontsize{10.2}{10}\selectfont
\begin{algorithmic}[1]
\STATE Initialization.
\FOR {$\textit{episode} = 1, \cdots, \textit{MAX}$}
\STATE The environment initializes the state $\boldsymbol s^1$ and agent initializes its observation $\boldsymbol{o}_m^1$.
\WHILE {the episode is not terminated}
\STATE For each agent $m \in \{1,\ldots,M\}$, sample action $\boldsymbol{a}_m^t$ according to the policy $\pi_{\theta}$.
\STATE Execute the joint action $\mathbf{a}^t=\{\mathbf{a}_m^t\}_{m=1}^M$.
\STATE Solve $\mathcal{P}2.1$ for each scheduled directed link $(m,m')\in\mathcal{E}$ to check ISAC QoS feasibility and obtain $R_q[t]$.
\STATE Compute the topology penalty based on SDR feasibility, and obtain $r^t$, $\mathbf{s}^{t+1}$, and $\{\mathbf{o}_m^{t+1}\}_{m=1}^M$.
\STATE Store the state transition in the rollout buffer $\mathcal D$.
\STATE Update $\boldsymbol{s}^t \leftarrow \boldsymbol{s}^{t+1}$ and $\boldsymbol{o}_m^t \leftarrow \boldsymbol{o}_m^{t+1}, \forall m$.
\ENDWHILE
\IF{Rollout buffer full}
\STATE Sample a random mini-batch of transitions from $\mathcal D$.
\STATE Update $\nu$ by computing $\nabla_{\nu}J_{a}(\nu)$ according to (\ref{au}) for agent $m \in \{1,\ldots,M\}$.
\STATE Update $\omega$ by computing $\nabla_{\omega}J_{c}(\omega)$ according to (\ref{cu}).
\ENDIF
\ENDFOR
\end{algorithmic}
}
\end{algorithm}

\vspace{-2mm}
\section{Simulation Results}
We consider a \(2.5\,\mathrm{km}\times 2.5\,\mathrm{km}\) power line corridor with a set of MDs.
The UAVs are assumed to fly at a fixed altitude of \(80\,\mathrm{m}\), with the initial position set to \((0,2500)\) and the destination located at \((2500,0)\) under a coordinate system. 
The simulation horizon is set to $T=500\,\mathrm{s}$, with the 5 min DLR update timescale \cite{poli2019possible} used as the timing reference.
Completing data collection within this window with lower UAV energy consumption could enable timely line-rating updates and support efficient grid operation.
For sensing QoS, the following three directions are considered, with central angles \([{-10}^{\circ},0^{\circ},10^{\circ}]\) and $\Gamma = -4$ dB.
The UAV propulsion energy parameters are adopted from \cite{8663615}, while the remaining system parameters and training hyperparameters are summarized in Table~I.

\begin{figure}[!ht]
   \centering
   \setlength{\abovecaptionskip}{2pt}
   \setlength{\belowcaptionskip}{-4pt}
   \includegraphics[width=3.5in]{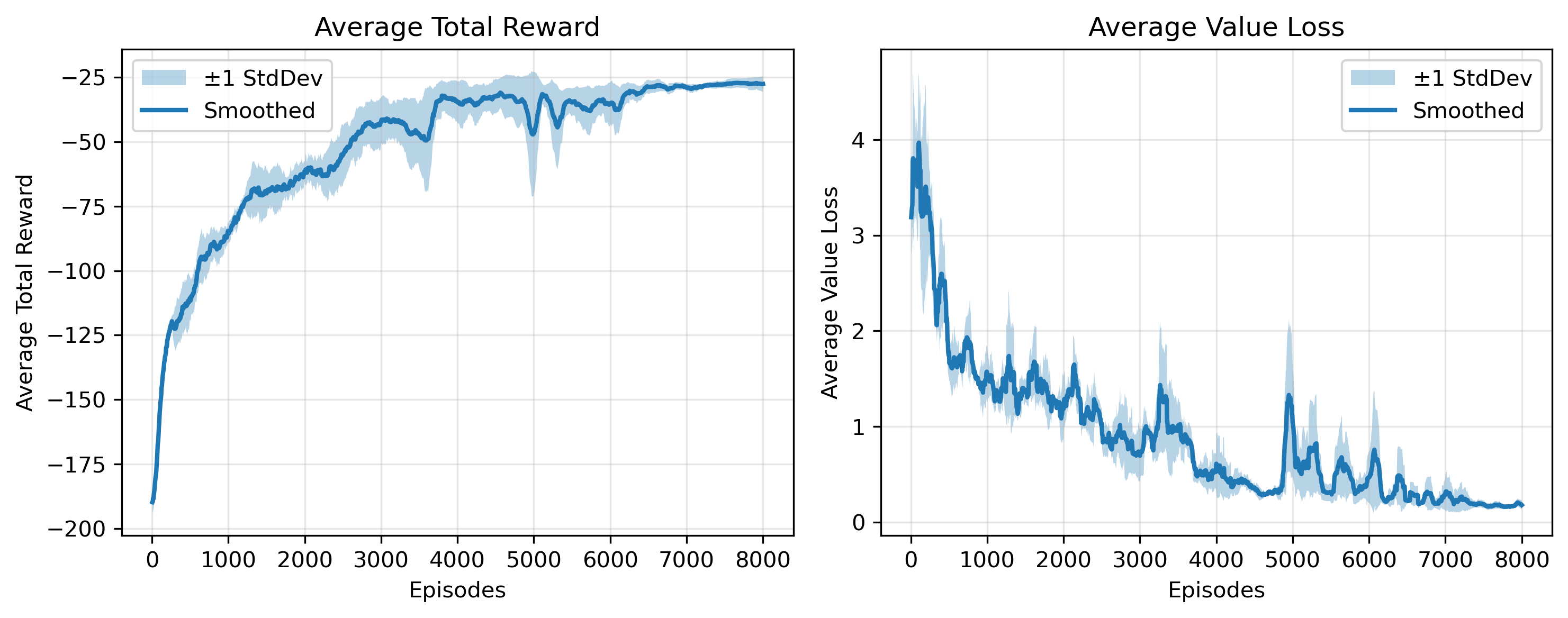}
   \vspace{-7mm}
   \caption{Training convergence of the proposed DRL-SDR algorithm over five runs, with reward and value loss smoothed every 20 episodes per run and then averaged across runs. The stabilized reward and value loss indicate that the algorithm converges during training.}
   \label{fig2}
   \vspace{-3mm}
\end{figure}

The convergence performance of the proposed DRL-SDR algorithm is shown in Fig.~\ref{fig2}. Next, five baselines are considered for comparison, as summarized in Table~\ref{baselines}.
The offline baselines retain only the sensing requirement, without optimizing inter-UAV communication beamforming.
The DRL with separate sensing-and-communication (DRL-S\&C) baseline uses an independent radar module in addition to the considered ULA, resulting in extra payload weight and an additional circuit power consumption of 2\,W~\cite{malle2021survey}.

Table~\ref{tab-uav} compares the energy and time consumption under different numbers of UAVs, with 30 MDs and $\gamma_{\text{th}}^{\text{U}} = 8$\,dB. 
The results show that multi-UAV cooperation substantially reduces the mission time, from about 6 minutes with one UAV to nearly half with multiple UAVs, thus satisfying the time requirement of DLR. 
Moreover, inter-UAV communication is especially beneficial when the number of UAVs is small, e.g., 2--4, as the proposed method achieves the lowest time and energy consumption.
\vspace{-2mm}
\begin{table}[!htbp]
\centering
\caption{System Parameters and DRL Hyperparameters}
\vspace{-2mm}
\label{tab:parameters}
\scriptsize
\setlength{\tabcolsep}{3pt}
\renewcommand{\arraystretch}{0.9}
\begin{tabular}{lc|lc}
\hline
\multicolumn{4}{c}{\textbf{System Parameters}} \\
\hline
Parameter & Value & Parameter & Value \\
\hline
Number of towers \& lines & 20, 20 
& ULA antennas per UAV & 12 \\

UAV safe distance ($d_{\min}$) & 10 m 
& UAV flying speed ($V_{\text{fixed}}$) & 20 m/s \\

MD transmit power ($p_i$) & 5 mW 
& Inter-UAV noise power ($\sigma_U^2$) & -94 dBm \\

UAV max power ($p_{\max}$) & 100 mW 
& MD-UAV noise power ($\sigma_{\text{MD}}^2$) & -101 dBm \\

TBP gain threshold ($\Gamma$) & -4 dB 
& MD-UAV SINR req. ($\gamma_{\text{th}}^{\text{MD}}$) & 3 dB \\
\hline
\multicolumn{4}{c}{\textbf{DRL Hyperparameters}} \\
\hline
Hyperparameter & Value & Hyperparameter & Value \\
\hline
Hidden layer dimension & 256 
& Discount factor ($\eta$) & 0.99 \\

Actor learning rate & $1\times10^{-4}$ 
& GAE parameter ($\lambda$) & 0.95 \\

Critic learning rate & $3\times10^{-4}$ 
& Clipping ratio ($\epsilon$) & 0.2 \\

Entropy coefficient & 0.01 
& Minibatch size & 256 \\
\hline
\end{tabular}
\vspace{-2mm}
\end{table}
\vspace{-3mm}

\begin{table}[!t]
\centering
\caption{Comparison schemes}
\vspace{-2mm}
\label{baselines}
\scriptsize
\setlength{\tabcolsep}{2.2pt}
\renewcommand{\arraystretch}{0.9}
\begin{tabular}{lcccc}
\hline
\textbf{Scheme} & \textbf{Method} & \textbf{Optimized Variables} & \textbf{Inter-UAV} \\
\hline
Proposed DRL-SDR & DRL \& SDR & $\boldsymbol{q}, \boldsymbol{b}, \boldsymbol{W}^c, \boldsymbol{R}^s$ & Connected \\
Greedy-online    & Greedy \& SDR  & $\boldsymbol{q}, \boldsymbol{b}, \boldsymbol{W}^c, \boldsymbol{R}^s$ & Connected \\
Greedy-offline   & Greedy  & $\boldsymbol{q}, \boldsymbol{b}$       & Disconnected \\
PSO-offline      & Particle swarm     & $\boldsymbol{q}, \boldsymbol{b}$  & Disconnected \\
GA-offline       & Genetic      & $\boldsymbol{q}, \boldsymbol{b}$        & Disconnected \\
DRL-S\&C         & DRL     & $\boldsymbol{q}, \boldsymbol{b}$        & Connected \\
\hline
\end{tabular}
\vspace{-2mm}
\end{table}

\begin{table}[!t]
\centering
\caption{Performance comparison under different numbers of UAVs.}
\vspace{-2mm}
\label{tab-uav}
\scriptsize
\setlength{\tabcolsep}{3.0pt}
\renewcommand{\arraystretch}{0.9}
\resizebox{\columnwidth}{!}{
\begin{tabular}{llccccc}
\hline
\textbf{Method} & \textbf{Metric} & \textbf{UAV=1} & \textbf{UAV=2} & \textbf{UAV=3} & \textbf{UAV=4} & \textbf{UAV=5} \\
\hline
\multirow{2}{*}{Proposed DRL-SDR}
& Energy ($\times 10^5$J) & $\mathbf{0.56}$ & $\mathbf{0.69}$ & $\mathbf{0.97}$ & $\mathbf{1.31}$ & 1.67 \\
& Total time (s) & 352 & 199 & 186 & 198 & 210 \\
\hline
\multirow{2}{*}{Greedy-offline}
& Energy ($\times 10^5$J) & 0.64 & 0.85 & 1.06 & 1.37 & $\mathbf{1.62}$ \\
& Total time (s) & 374 & 268 & 229 & 216 & 202 \\
\hline
\multirow{2}{*}{Greedy-online}
& Energy ($\times 10^5$J) & 0.64 & 0.71 & 1.02 & 1.38 & 1.71 \\
& Total time (s) & 374 & 219 & 218 & 208 & 204 \\
\hline
\multirow{2}{*}{PSO-offline}
& Energy ($\times 10^5$J) & 0.61 & 0.95 & 1.17 & 1.46 & 1.68 \\
& Total time (s) & 353 & 286 & 280 & 237 & 221 \\
\hline
\multirow{2}{*}{GA-offline}
& Energy ($\times 10^5$J) & 0.69 & 1.03 & 1.19 & 1.41 & 1.70 \\
& Total time (s) & 400 & 374 & 249 & 263 & 220 \\
\hline
\multirow{2}{*}{DRL-S\&C}
& Energy ($\times 10^5$J) & 0.57 & 0.71 & 0.99 & 1.34 & 1.71 \\
& Total time (s) & 352 & 199 & 186 & 198 & 210 \\
\hline
\end{tabular}
}
\vspace{-3mm}
\end{table}

\begin{figure}[!t]
   \centering
   \setlength{\abovecaptionskip}{2pt}
   \setlength{\belowcaptionskip}{-4pt}
   \includegraphics[width=2.6in]{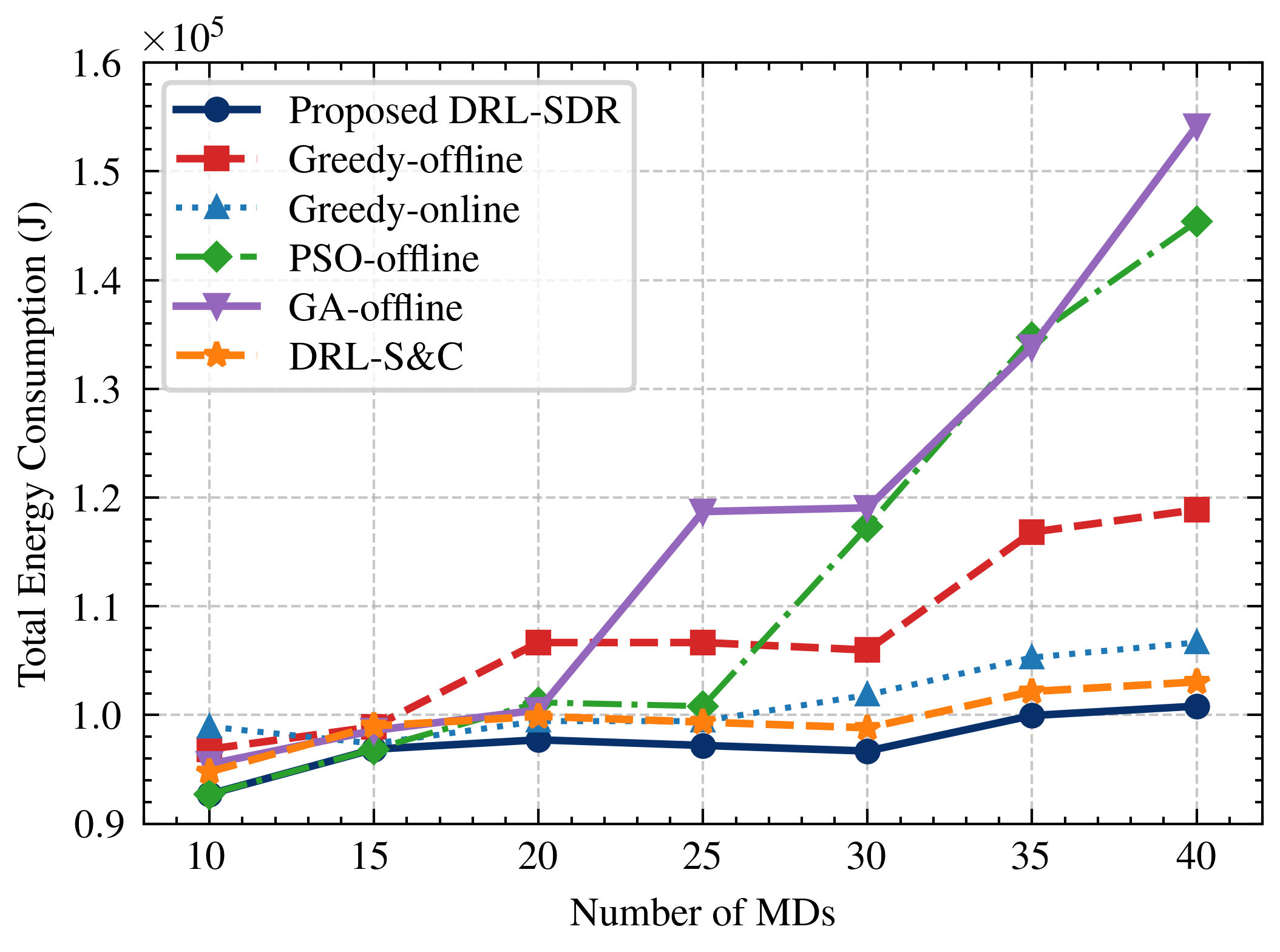}
   \vspace{-3mm}
   \caption{Total energy consumption vs. MD number.}
   \label{fig:md_compare}
   \vspace{-3mm}
\end{figure}
Fig.~\ref{fig:md_compare} investigates the effect of MD number on total energy consumption, with 3 UAVs and $\gamma_{\text{th}}^{\text{U}} = 8$\,dB. 
It can be seen that the schemes with inter-UAV connection generally maintain lower and more stable energy consumption as the number of MDs increases, which indicates the benefit of inter-UAV cooperation under heavier collection demand. 
Among all schemes, the proposed DRL-SDR achieves the lowest energy consumption. 
Specifically, at MD $=40$, the proposed DRL-SDR reduces energy consumption by 34.60\%, 30.65\%, 15.20\%, 5.48\%, and 2.16\% compared with GA, PSO, Greedy-offline, Greedy-online, and DRL-S\&C, respectively.

\begin{figure}[!t]
   \centering
   \setlength{\abovecaptionskip}{2pt}
   \setlength{\belowcaptionskip}{-4pt}
   \includegraphics[width=2.6in]{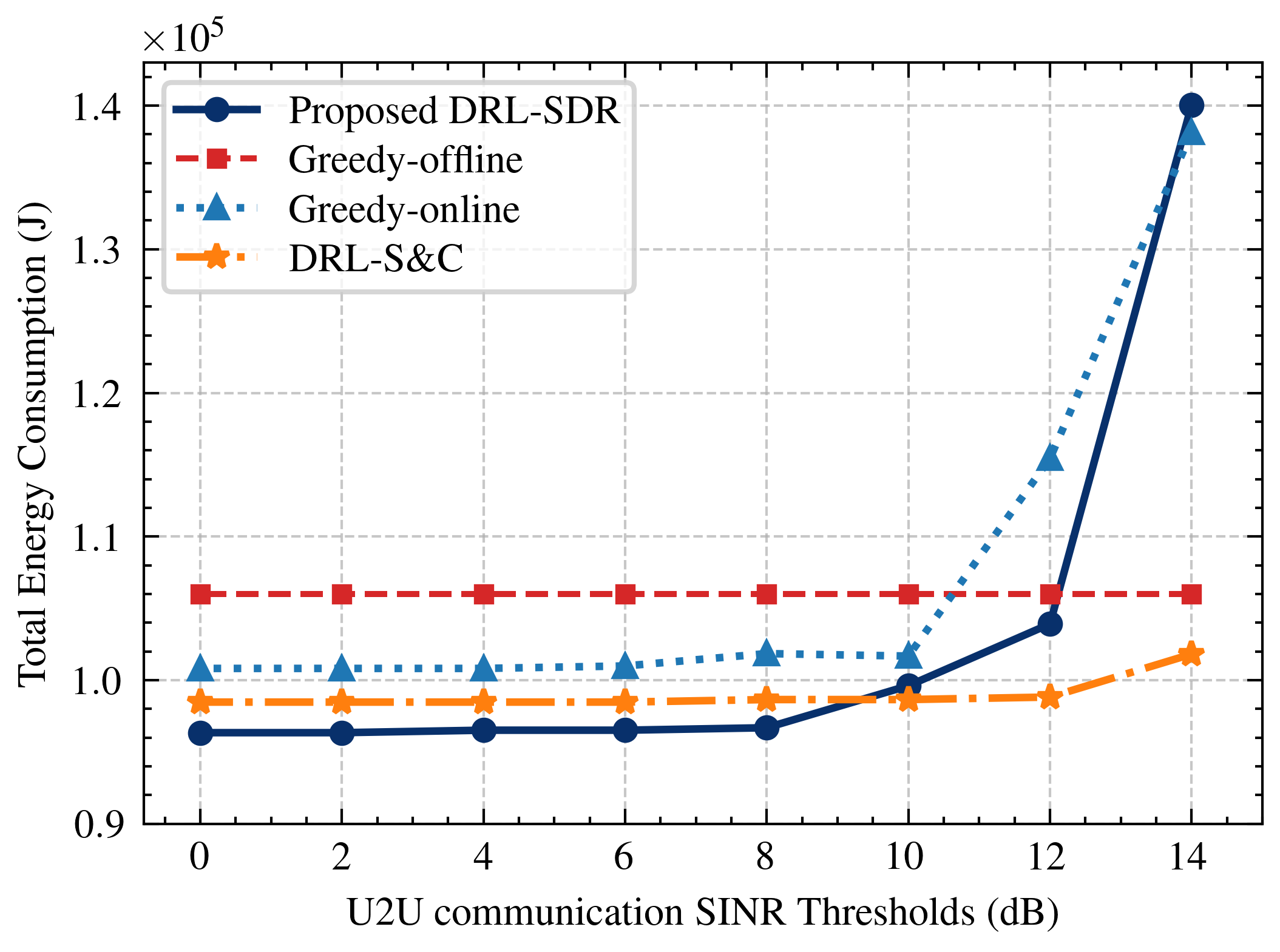}
   \vspace{-3mm}
   \caption{Total energy consumption vs. Inter-UAV communication SINR.}
   \label{fig:sinr_compare}
   \vspace{-3mm}
\end{figure}
In Fig.~\ref{fig:sinr_compare}, the effect of different inter-UAV communication SINR thresholds $\gamma_{\text{th}}^{\text{U}}$ is evaluated with 3 UAVs and 30 MDs under the given sensing QoS (i.e., $\Gamma = -4$ dB).
For most SINR thresholds, DRL-S\&C consumes about 2.2\% more energy than the proposed DRL-SDR scheme, demonstrating the energy benefit of ISAC. 
When $\gamma_{\mathrm{th}}^{\mathrm{U}}$ increases over 10 dB, DRL-S\&C outperforms the proposed method, since the stringent inter-UAV communication requirement increases the ISAC beamforming cost. 
At 14 dB, Greedy-offline also consumes less energy than DRL-SDR, showing that strict communication requirements could offset the benefit of UAV cooperation.

\vspace{-2mm}
\section{Conclusion}
In this paper, we study an ISAC-enabled UAV-based data collection framework for timely DLR monitoring in smart grids.
Specifically, we formulate a joint energy minimization problem for collection scheduling and UAV trajectory optimization subject to ISAC QoS constraints, and solve it using a hierarchical DRL-SDR algorithm.
Simulation results show that the proposed algorithm reduces total energy consumption compared with the baselines while meeting the minute-level DLR requirement. 
The results also confirm the benefit of inter-UAV connectivity for small UAV groups and show that the proposed scheme maintains the best energy efficiency for inter-UAV SINR thresholds up to 8 dB under the considered sensing QoS constraints.
Future work will extend the proposed framework to ISAC-assisted hazard-aware 3D trajectory design in power line corridors, where sensing feedback is used to support safe UAV maneuvering while maintaining efficient DLR data collection.

\vspace{-2mm}
\normalem
\begingroup
\fontsize{8}{8.0}\selectfont
\bibliographystyle{IEEEtran}
\bibliography{ref}

@ARTICLE{8663615,
  author={Zeng, Yong and Xu, Jie and Zhang, Rui},
  journal={IEEE Transactions on Wireless Communications},
  title={Energy Minimization for Wireless Communication With Rotary-Wing UAV},
  year={2019},
  volume={18},
  number={4},
  pages={2329-2345},
  doi={10.1109/TWC.2019.2902559}}

@ARTICLE{10791445,
  author={Wang, Yi and Zu, Keke and Xiang, Luping and Zhang, Qixun and Feng, Zhiyong and Hu, Jie and Yang, Kun},
  journal={IEEE Transactions on Wireless Communications},
  title={ISAC Enabled Cooperative Detection for Cellular-Connected UAV Network},
  year={2025},
  volume={24},
  number={2},
  pages={1541-1554},
  doi={10.1109/TWC.2024.3509978}}

@ARTICLE{10769423,
  author={Liu, Yaxi and Mao, Wencan and He, Boxin and Huangfu, Wei and Huang, Tianyao and Zhang, Haijun and Long, Keping},
  journal={IEEE Transactions on Communications},
  title={Radar Probing Optimization for Joint Beamforming and UAV Trajectory Design in UAV-Enabled Integrated Sensing and Communication},
  year={2025},
  volume={73},
  number={6},
  pages={4469-4485},
  doi={10.1109/TCOMM.2024.3506917}}

@ARTICLE{10603401,
  author={He, Keren and Zhou, Quan and Lian, Zhijie and Shen, Yang and Gao, Jiachen and Shuai, Zhikang},
  journal={IEEE Transactions on Industry Applications},
  title={Spatiotemporal Precise Routing Strategy for Multi-UAV-Based Power Line Inspection With Integrated Satellite-Terrestrial Network},
  year={2024},
  volume={60},
  number={6},
  pages={8418-8429},
  doi={10.1109/TIA.2024.3430248}}

@article{karimi2018dynamic,
  title={Dynamic thermal rating of transmission lines: A review},
  author={Karimi, Soheila and Musilek, Petr and Knight, Andrew M},
  journal={Renewable and Sustainable Energy Reviews},
  volume={91},
  pages={600--612},
  year={2018},
  publisher={Elsevier}
}

@article{poli2019possible,
  title={The possible impact of weather uncertainty on the Dynamic Thermal Rating of transmission power lines: A Monte Carlo error-based approach},
  author={Poli, Davide and Pelacchi, Paolo and Lutzemberger, Giovanni and Scirocco, Temistocle Baffa and Bassi, Fabio and Bruno, Gianluca},
  journal={Electric Power Systems Research},
  volume={170},
  pages={338--347},
  year={2019},
  publisher={Elsevier}
}

@article{dianovsky2023electromagnetic,
  title={Electromagnetic radiation from high-voltage transmission lines: Impact on uav flight safety and performance},
  author={Dianovsk{\`y}, Robert and Pecho, Pavol and Vel'k{\`y}, Patrik and Hr{\'u}z, Michal},
  journal={Transportation research procedia},
  volume={75},
  pages={209--218},
  year={2023},
  publisher={Elsevier}
}

@ARTICLE{10097711,
  author={Li, Li and Li, Wei and Wang, Jun and Chen, Xiaonan and Peng, Qihang and Huang, Wei},
  journal={IEEE Communications Letters}, 
  title={UAV Trajectory Optimization for Spectrum Cartography: A PPO Approach}, 
  year={2023},
  volume={27},
  number={6},
  pages={1575-1579},
  doi={10.1109/LCOMM.2023.3265214}}

@article{wydra2019time,
  title={Time-aware monitoring of overhead transmission line sag and temperature with LoRa communication},
  author={Wydra, Michal and Kubaczynski, Pawel and Mazur, Katarzyna and Ksiezopolski, Bogdan},
  journal={Energies},
  volume={12},
  number={3},
  pages={505},
  year={2019},
  publisher={MDPI}
}

@inproceedings{malle2021survey,
  title={Survey and evaluation of sensors for overhead cable detection using UAVs},
  author={Malle, Nicolaj Haarh{\o}j and Nyboe, Frederik Falk and Ebeid, Emad},
  booktitle={2021 International Conference on Unmanned Aircraft Systems (ICUAS)},
  pages={361--370},
  year={2021},
  organization={IEEE}
}

@ARTICLE{10065524,
  author={Wang, Xijun and Yi, Mengjie and Liu, Juan and Zhang, Yan and Wang, Meng and Bai, Bo},
  journal={IEEE Transactions on Communications}, 
  title={Cooperative Data Collection With Multiple UAVs for Information Freshness in the Internet of Things}, 
  year={2023},
  volume={71},
  number={5},
  pages={2740-2755},
  doi={10.1109/TCOMM.2023.3255240}}

@ARTICLE{10879807,
  author={Cheng, Gaoyuan and Song, Xianxin and Lyu, Zhonghao and Xu, Jie},
  journal={IEEE Transactions on Communications}, 
  title={Networked ISAC for Low-Altitude Economy: Coordinated Transmit Beamforming and UAV Trajectory Design}, 
  year={2025},
  volume={73},
  number={8},
  pages={5832-5847},
  doi={10.1109/TCOMM.2025.3541027}}
\endgroup
\end{document}